\documentclass[preprint,12pt]{elsarticle}

\usepackage{amssymb}
\usepackage{float}
\usepackage{multirow}

\journal{Cement and Concrete Research}

\begin{document}

\begin{frontmatter}



\title{Ultrasonic monitoring of carbonation in Portland cements: non linear and linear analysis}

\author[label1,label2]{A. Villarreal}
\author[label1]{P.F.J. Cano-Barrita}
\author[label1]{F.M. Leon-Martinez}
\author[label2]{L. Medina}
\author[label1]{F. Castellanos}\ead{fcastellanos@ipn.mx}
\affiliation[label1]{organization={Instituto Politécnico Nacional-CIIDIR, Unidad Oaxaca},
             addressline={C. Hornos 1003},
             city={Santa Cruz Xoxocotlán},
             postcode={C.P. 71233},
             state={Oaxaca},
             country={México}}
\affiliation[label2]{organization={Universidad Nacional Autónoma de México, Facultad de Ciencias},
             addressline={Ciudad Universitaria},
             city={Coyoacan},
             postcode={C.P. 04510},
             state={CDMX},
             country={México}}

\begin{abstract}
Chemical reactions resulting from the ingress of carbonates into the cement matrix modify the properties of its pore solution, as well as its pore distribution and size. These changes lead to corrosion of the steel in reinforced concrete. The nature of conventional testing for the estimation of carbonation in cement-based materials is time-consuming and destructive. This paper presents a set of non-destructive ultrasound-based indexes, obtained solely from non-linear and linear analyses of ultrasonic signals, for measuring the carbonation of Portland cement pastes. Class 30RS cement pastes with three water/cement ratios by weight (0.4, 0.5, and 0.6) were considered. Carbonation was carried out for 120 days with a constant CO2 level of 4
\end{abstract}

\begin{keyword}
Ultrasound \sep Carbonation \sep Elastic Moduli \sep Microstructure

\end{keyword}

\end{frontmatter}


\section{INTRODUCTION}\label{intro}
Carbonation is a chemical process that occurs when atmospheric carbon dioxide $(CO_2)$ dissolved in a pore solution forms carbonic acid that reacts with calcium hydroxide $(Ca(OH)_2)$ to form calcium carbonate $(CaCO_3)$ and water $(H_2O)$. Carbonation results in a decrease in the alkalinity of the pore solution, changes in pore size distribution, and a decrease in total porosity. In reinforced concrete, the decreased alkalinity at the level of the reinforcing steel leads to its depassivation, and a corrosion process may start.
Changes produced by the carbonation of cement paste are manifested in its physical properties, such as density, Young’s modulus, and shear modulus. These changes modify the linear elastic behavior of the material \cite{Yang}.
Carbonation mainly influences the behavior of cement paste in two ways:
\begin{enumerate}
\item Hardening effect: Carbonation can increase the strength and rigidity (stiffness) of cement paste through the formation of $CaCO_3$ crystals within the pore structure. These crystals reduce pore size because the molar volume of $CaCO_3$ is 11\%-14\% higher than that of the reacted calcium hydroxide \cite{Xi,Morandeau}, which leads to a denser microstructure. The increase in rigidity results in a more linear stress–strain response.
\item Chemical composition changes: Carbonation alters the chemical composition of cement paste when consuming calcium hydroxide to form $CaCO_3$, especially in the pore solution. $CaCO_3$ is a less soluble compound than calcium hydroxide and relatively stable. The decrease in calcium hydroxide content decreases the pH of the pore solution, affects the chemical interactions within the cement paste, and changes the mechanical behavior of the material. 
\end{enumerate}
Linear analysis of traditional ultrasonic tests have limitations in assessing the microstructure of cement pastes \cite{Chen}. These tests include phase velocity analysis and/or attenuation. Although these techniques can identify changes in the microstructure of cement paste, sometimes this identification is not evident. Therefore, it is necessary to develop more precise and efficient techniques for characterizing the existence of micro-cracks or modifications in the microstructure of cement-based materials.
In most linear analysis, the presence of damage is assessed by comparing the received wave signal with a benchmark. However, linear analysis of waves has limitations in distinguishing changes in the response signal caused by damage from those originating from other sources, such as carbonation. In contrast, non-linear ultrasonic techniques have exhibited greater sensitivity to low-level damage in comparison to linear analysis \cite{Amerini}; moreover, they do not depend on any benchmark. These techniques analyze high-frequency harmonics and subharmonics resulting from the non-linear interactions between ultrasonic waves and the material, that is, harmonics at frequencies that are multiples of the fundamental frequency of the incident wave, which will depend on the characteristics of the media. As the increase in this non-linear interaction is directly proportional to the damage, the extent of damage can be determined by measuring the non-linearity of the ultrasonic wave propagation through the material. Thus, non-linear analysis complements and enhances traditional ultrasonic techniques \cite{Drai,Abbate,Huang}. 

The classical theory of non-linear acoustics states that second harmonics are mainly induced by the characteristics of the microstructure of materials \cite{Matlack,Suzuki,Hikata}, whereas the effect of “non-classical” non-linearity is caused by microcracks in materials \cite{Ostrovsky}; generally, the effect of the latter is considered stronger than that of the former. Non-classical, non-linear models mainly include the contact surface model, the hysteresis model, and the bilinear stiffness model \cite{Broda}. Based on these models, Zhao et al. \cite{Zhao} derived an analytical expression of the non-linear parameters and elastic moduli of solids and verified such a theoretical model with numerical simulations. Nam et al. \cite{Nam} proposed a theoretical model for the evaluation of localized microdamage on both the internal and external surface of the structure of aluminum 6061-T6, according to the amplitude of the second harmonic. 

To explore the non-linear ultrasonic response in cement pastes under different loading patterns, Ongpeng et al. \cite{Ongpenq} conducted non-linear ultrasonic tests in loading and unloading situations in cement paste specimens with different water/cement ratios. The results showed that third-order harmonics are more sensitive to the uni axial compression pattern, whereas second-order harmonics are more sensitive to the multiple loading and unloading pattern. In general, non-linear ultrasonic techniques have become one of the most powerful methods for the initial detection of damage in materials \cite{Kim,Ongpeng2,Hirose}.

Non-linear techniques can be used to quantify cement carbonation, as a carbonated sample exhibits a decrease in porosity caused by a greater molar volume of carbonation products compared to the volume of hydrates. For example, the molar volume of $Ca(OH)_2$ is 33.2 $cm^3/mol$, whereas that of $CaCO_3$ is 36.9 $cm^3/mol$, corresponding to an 11\% increase. The impact of this variation can be seen in cement pastes and concrete using mercury intrusion porosimetry \cite{Shah,Thiery}. The difference in porosity between carbonated and non-carbonated concrete, which can reach 10\%, is higher when the water/cement (w/c) ratio increases \cite{Thiery}. The same studies emphasize that carbonation also alters pore size distribution, with an increase in pore size commonly observed after carbonation. For example, an increase in the volume of capillary pores was observed after complete carbonation in cement pastes prepared with blended cements \cite{Shah}. These authors also comment that coarsening of the pore structure may be associated with additional silica gel formation due to the decomposition of $C-S-H$ after a prolonged exposure to $CO_2$. For Portland cement, Bouchaale et al. \cite{Bouchaala} demonstrated that standard mechanical properties such as the modulus of elasticity increase with carbonation. This change results in an increase in wave group velocity of approximately 2\%. Recent research has demonstrated the influence of water saturation and alkali silica reaction on changes in porosity \cite{Payan,Chen2}. Nevertheless, there have been few studies specifically exploring the influence of carbonation on the non-linear behavior of concrete. For example, Gross et al. \cite{Gross} used Rayleigh waves to characterize carbonation in cement-based materials.  

In summary, numerous studies and techniques have been developed to analyze the non-linearity of cement pastes. These studies have highlighted the sensitivity of these pastes to damage in different scenarios and application fields. However, very few studies have used nondestructive ultrasonic testing to analyze signals and assess changes in the linear properties of cement pastes due to carbonation \cite{Kim,Zhao2,Bui}.

In this study, linear and non-linear analyses of ultrasonic responses of carbonated samples are proposed to correlate changes in ultrasonic signals with those in the microstructure of cement pastes due to carbonation. The proposed study aims to investigate the energy of ultrasonic signals as a means of examining attenuation mechanisms due to wave interaction with the carbonated microstructure. Furthermore, analyzing the signal phase provides information on the dispersion of the signal. Additionally, non-linear analysis of the harmonics amplitude produced by power spectral analysis is performed, in which a defined non-linear parameter correlates with the level of carbonation of the microstructure. 

\section{Methodology (or materials and method)}
\subsection{Materials}
Ordinary Portland cement class 30RS was used to prepare cement pastes with w/c ratios of 0.4, 0.5, and 0.6 by weight, according to ASTM C305-99 standard \cite{ASTM}. Twenty-four cylindrical specimens were cast from each w/c ratio. The specimens were cured in a saturated calcium hydroxide solution at $40\pm1$ $^oC$ for 56 days to avoid cement paste leaching, which would lead to an increase in porosity and decrease in density, as well as a change in the modulus of elasticity. The dimensions of each specimen were $50\pm2$ mm in diameter by $57\pm2$ mm in height, after correcting for obtaining parallel and smooth cylinder surfaces. Once the cylinders were ready, they were sanded (P-180), and the dust was removed using distilled water. Later, epoxy resin was applied to all surfaces except for both bases of the cylinder at each end. 

Prior to exposing the samples to the carbonation process, they were equilibrated at $30\pm1$ $^oC$ and $65\pm1$\% relative humidity until their minimum change in weight was $<0.5$\% of the registered values in a 24-h interval. Small cracks were generated due to drying shrinkage of the samples, which were observed mainly in the 0.60 w/c ratio samples.

Once the equilibrium conditions were reached, the specimens were carbonated at a constant $CO_2$ concentration of $4\pm0.1$\% by volume, considering a unidirectional carbonation parallel to the longitudinal axis. One of the bases of the cylinder was protected using parafilm. Two samples of each w/c ratio were kept as controls within an environmental chamber with the same ambient conditions but without $CO_2$. The samples were kept in the chambers for 120 days and taken out at 1, 7, 14, 28, 56, and 120 days for ultrasonic wave acquisition (approximately 1 h for testing). 

\subsection{FTIR measurements}
Specimens of each carbonated and non-carbonated cement paste were taken at various time intervals from 1 to 56 days to perform Fourier-transform infrared spectroscopy (FTIR) measurements \cite{Cosmes}. The cement paste samples were prepared by crushing and grinding with an agate mortar and pestle after removing the epoxy resin layer. The resulting fine powder was sieved and oven-dried at $105\pm1$ $^oC$ before being stored in sealed bags with desiccants to prevent moisture adsorption and carbonation. FTIR spectra of the powder samples were obtained using a mid-range Nicolet 6700 FTIR spectrometer (Thermo Scientific, USA) with a single bounce Smart iTR diamond ATR accessory. The area under the curve of the vibration bands corresponding to the carbonate ion in $CaCO_3$ was calculated. The initial amount of $CaCO_3$ in the cement paste was determined by thermogravimetric analysis and used as a reference to estimate the amount of $CaCO_3$ by weight in the carbonated specimens \cite{Zhu}. 

\subsection{Signal acquisition}
\begin{figure}[h]
\centering
\includegraphics[width=1\linewidth]{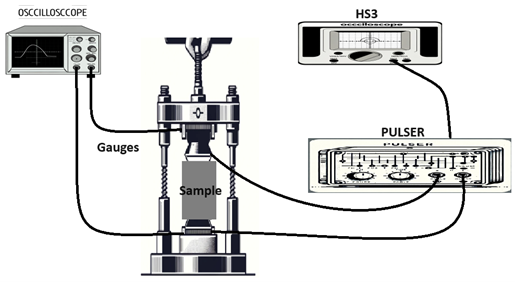}\caption{Diagram of the signal acquisition system\cite{Villarreal}}\label{fig:esq}
\end{figure}

Through-transmission ultrasonic signals were acquired for 18 specimens (6 of each w/c ratio) at different exposure times to carbonation (0, 1, 3, 5, 7, 14, 28, 56, and 120 days) to determine changes in their physical properties by identifying variations in the signals. The mean of the signals was calculated and employed for their analysis. The specimens were exposed to longitudinal ultrasonic waves generated by transducers with frequencies of 500 kHz (Olympus, Tokyo, Japan). The orientation of the transducers and the contact pressure were kept constant during the experiment using a base and pressure transducers (Figure \ref{fig:esq}). Petroleum jelly was used as a couplant to improve wave transmission. The excitation signal (approximating a Kronecker delta) was generated using a 5058PR emitter/receiver (Olympus, Tokyo, Japan) with a voltage of 200 V. An amplification of 40 dB was used for signal acquisition. 
 
\subsection{Signal processing}
\subsubsection{Energy of a signal}
Energy can be defined as a power measure contained in the signal over time. For discrete and digital signals, energy can be calculated using Equation 1:
\begin{equation}
\textrm{Discrete signal energy} = \Sigma x[n]^2
\end{equation}
Where $x[n]$ is the value of the signal in the discrete time instant $n$.

The term $x^2$ is the square of the signal at each instant of time. The total energy of the signal is obtained by discrete integrating these square values over the duration of the signal. 

For the case of voltage signals, a definition of the energy can be scaled using a factor of $1/Z$ (Equation 2):  

\begin{equation}
E=\frac{E_x}{Z}=\frac{1}{Z}\int_{-\infty}^{\infty}|x(t)|^2\,dt
\end{equation}
where Z is the impedance.

\subsubsection{Frequency spectrum}
Power spectral density is a fundamental concept in signal processing that provides information regarding the power distribution of a signal in its different frequency components. It is commonly used in various applications, such as the analysis of noise characteristics, noise power estimation, characterization of systems, and the study of physiological signals \cite{Smith}. The power spectral density of an ultrasonic signal is estimated by the discrete Fourier transformation (DFT) \cite{Kim2}. In this case, the DFT is used to analyze frequency components of the ultrasonic signal \cite{Thiery,Noguchi}. 

The phase of the signal is calculated from the DFT of the signal \cite{Winkler}. The analysis considers a continuous phase, without singularities, in multiples of the angle $2\pi$ radians, by “unwrapping/unfolding” the phase of the signals. This phase is employed for the estimation of the travel time of the ultrasonic wave in the specimen \cite{Kluk,Zhang,Peng}. 

\subsubsection{Non-linearity parameter}
According to linear acoustic theory, if a unidimensional longitudinal wave propagates in an isotropic medium and the amplitude of the elastic wave is small, then the stress remains linear. This means that changes in parameters such as the amplitude of the wave and wave velocity can be less sensitive to the changes in microstructure or porosity, as well as to microcracks in the material \cite{Zhao3}.
The ratio between the amplitudes of the fundamental harmonic and the following harmonics can be expressed as follows \cite{Zhao3}: 

\begin{equation}
\beta=\frac{8A_2}{A_1^2k^2x}\label{eq3}
\end{equation}

where $A_1$ and $A_2$ represent the amplitudes of the fundamental and the next observed harmonic, $x$ is the displacement, and $k$ the wave number. 
Equation \ref{eq3} can be rewritten as Equation \ref{eq4}, as the change of $\beta$ can be considered to only depend on the amplitude of the harmonics: 

\begin{equation}
\beta\approx\frac{A_2}{A_1^2}\label{eq4}
\end{equation}
Similarly, if this equation is extended to the third order, the non-linear parameter of the third order can be defined as:
\begin{equation}
\gamma\approx\frac{A_3}{A_1^3}\label{eq5}
\end{equation}
where $A_n$ stands for the nth harmonic amplitude. 
\section{Results and discussion}
\begin{figure}[H]
\centering
\includegraphics[width=8.5cm]{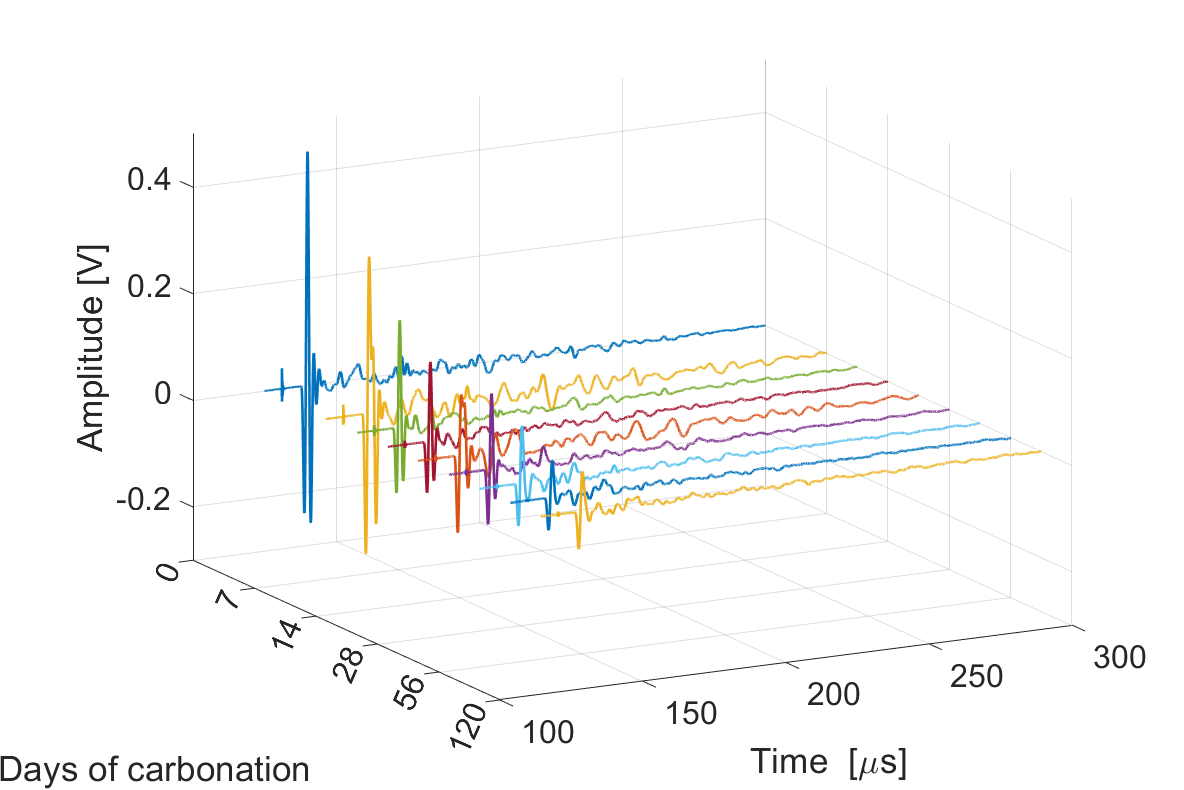}\caption{Ultrasonic signals of w/c = 0.6 specimens recorded after different days of exposure to $CO_2$.}\label{fig2}
\end{figure}

The following results are obtained through an acoustic analysis of the carbonation. As an example, Figure \ref{fig2} only shows the response signals of the specimens with a w/c ratio of 0.6, as all w/c ratios showed a similar behavior. The amplitude of the signal clearly decreased as the sample carbonated, which indicated energy losses of the transmitted wave, most likely due to changes in acoustic impedance caused by the carbonation front. 

\begin{figure}[H]
\centering
\includegraphics[width=0.9\linewidth]{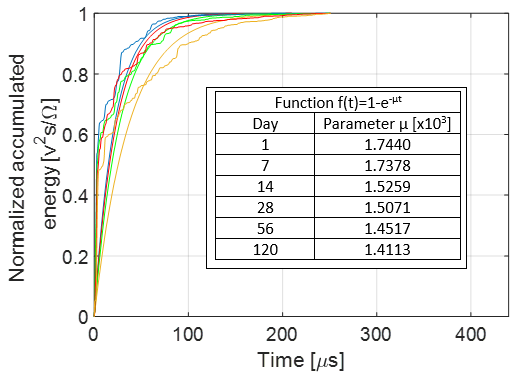}\caption{Carbonation exposure time vs. normalized cumulative energy of signals for the 0.6 w/c ratio samples. The color of the lines indicates the number of days of exposure to carbonation: 7 d (blue), 28 d (red), 56 d (green), and 120 d (yellow).}\label{fig3}
\end{figure}

Figure \ref{fig3} presents the results of the time-domain analysis of variations in energy caused by changes in the propagation medium. A normalized cumulative energy (Eq. 1) behavior is observed when the microstructure changes due to the carbonation. The energies are shown as carbonation increases. As the carbonation time increases, both the signal dispersion and the attenuation increase, as the signal energy becomes more scattered and takes longer to reach the normalized maximum. This behavior results from changes produced in the microstructure. The cumulative function $F(x)=1-e^{-\mu x}$ was fitted for the description of this cumulative energy using the parameter $\mu$ as an index of the carbonation, which is directly associated with the signal dispersion. The $\mu$ values of all pastes with different w/c ratios decreased as the carbonation in the specimens increased. For specimens with lower w/c ratios, this behavior was less evident, as the carbonation was not complete, and few dispersers were generated as a result of the carbonation itself.  

\begin{figure}[H]
\centering
\includegraphics[width=0.9\linewidth]{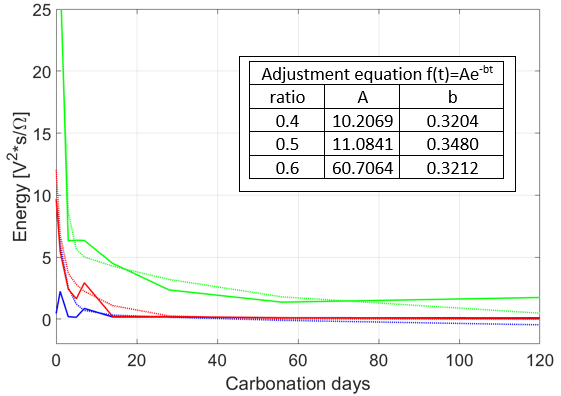}\caption{Carbonation time vs. total signal energy. The color of the lines indicates the w/c ratio: blue (w/c = 0.4), red (w/c = 0.5), and green (w/c = 0.6).}\label{fig4}
\end{figure}

This analysis can be complemented by the changes in the total energy of every signal over time as the carbonation progresses (Figure \ref{fig4}). The energy is calculated assuming a 1 $\Omega$ resistance to convert units of volts into joules (Eq. 2). As carbonation increased, a greater attenuation of the signal energy was observed. This behavior can be attributed to the fact that more rigid and dense pastes consume more energy per unit distance of propagation.

When the decreasing exponential function $Ae^{-bx}$ is used to quantify this energy decrease, the corresponding rate of decrease coefficient $b$ was $0.3204$, $0.3480$, and $0.3212$ for pastes with w/c ratios of 0.4, 0.5, and 0.6, respectively. These coefficients provide the energy decrease velocity for each carbonation time. The determination coefficients obtained were 0.95, 0.97, and 0.97, respectively. However, it was not possible to identify a trend related to the different w/c ratios.

\begin{figure}[H]
\centering
\includegraphics[width=0.9\linewidth]{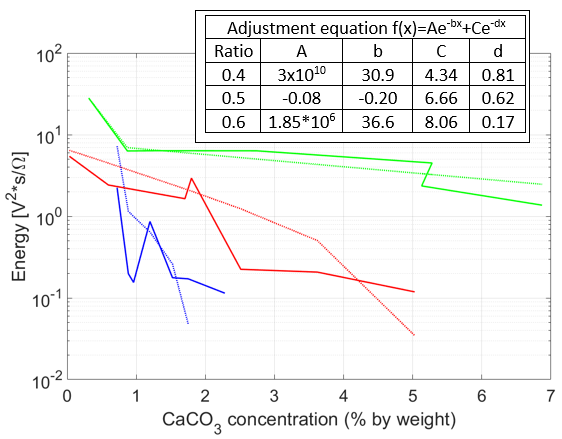}\caption{$CaCO_3$ concentration vs. total signal energy. The color of the lines indicates the w/c ratio: blue (w/c = 0.4), red (w/c = 0.5), and green (w/c = 0.6).}\label{fig5}
\end{figure}

Figure \ref{fig5} shows energy (Eq. 2) as a function of $CaCO_3$ concentration. It can be seen that the fastest change occurred in the signal that propagated in the specimen with a w/c ratio of 0.4, and the slowest change occurred in the specimen with a w/c ratio of 0.6. These results suggest that by increasing carbonation in the 0.6 w/c ratio specimen, a slower change in energy was obtained in comparison with the 0.4 w/c ratio specimen. A two-term exponential function $Ae^{-bx} + Ce^{-dx}$ was employed to fit the data, yielding the following determination coefficients (0.97, 0.85, and 0.98). It is considered that each exponential, and, therefore, each decay constant, can be associated with different loss energy processes related to carbonation. However, a description of these processes requires a more detailed analysis that is beyond the scope of this paper. 

\begin{figure}[H]
\centering
\includegraphics[width=0.9\linewidth]{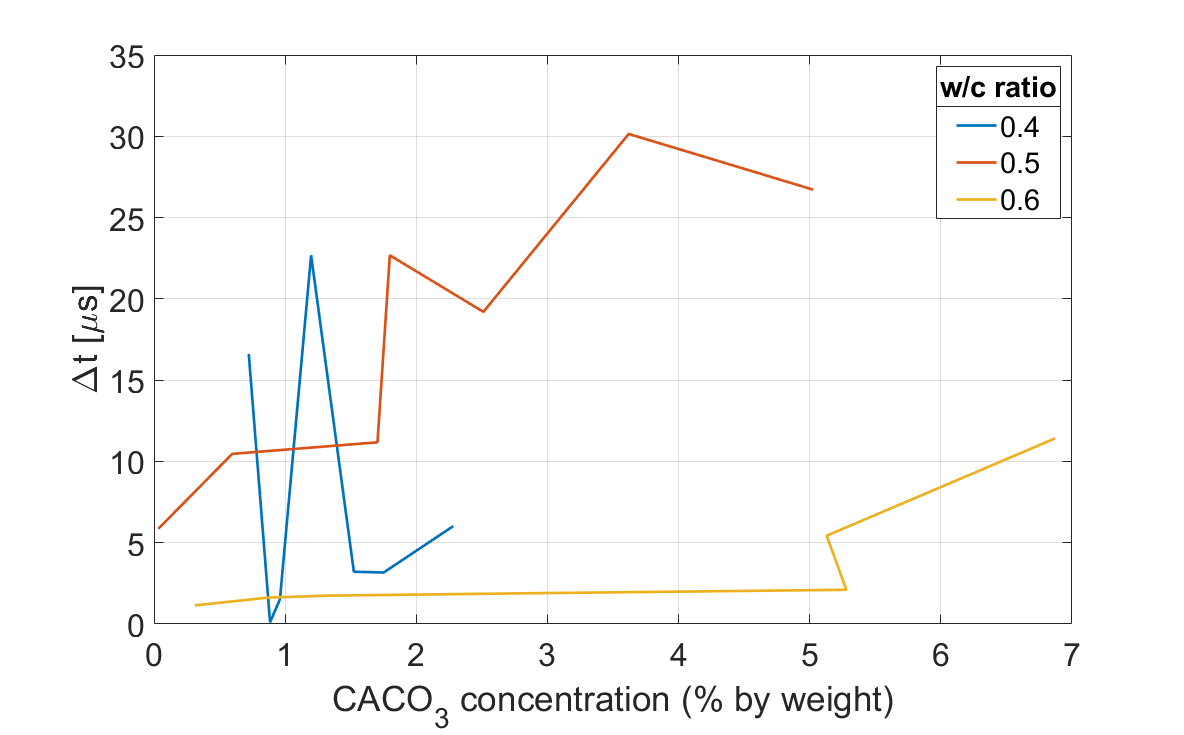}\caption{Duration $\Delta t$ as a function of $CaCO_3$ concentration.}\label{fig6}
\end{figure}

Further analysis of the cumulative normalized energy of the signal (Figure \ref{fig3}) allows us to observe the dispersion of energy as the signal propagates with increasing carbonation. For this analysis, the duration $\Delta t$ was calculated between the times when the cumulative normalized energy of the signal was 5\% and 95\% (Figure \ref{fig6}). From this duration, it is possible to analyze the dispersion behavior with respect to the $CaCO_3$ concentration. It was observed that the dispersion (evaluated through the standard deviation) was higher for the 0.5 w/c ratio specimen. The 0.6 w/c ratio specimen showed very little signal dispersion, where a change was observed only in the last days of the test. It is interesting to note that the 0.4 w/c ratio specimen did not show a clear trend, which must be related to its low carbonation. The coefficients of variation increased with increasing $CaCO_3$ concentration and w/c ratio (0.3038, 0.6623, and 0.9657 for 0.4, 0.5, and 0.6 w/c ratio specimens, respectively), which indicated that the energy, according to this measure, had a higher dispersion at higher w/c ratios.

For the linear analyses performed on an ultrasonic signal, we calculated the phase velocity and attenuation of the signal, among others. Attenuation was indirectly analyzed by means of the energy loss of the signal (Figures \ref{fig4} and \ref{fig5}). For the calculation of the velocity, it was necessary to obtain the phase behavior of a signal and compare it with a benchmark \cite{Villarreal}. A linear frequency-domain analysis of the ultrasonic signals, such as phase shift, was performed, which is related to the calculation of phase velocity. The phase of the signal propagating in a non-carbonated material was taken as a benchmark, so that the changes in the phases of the signals could serve as a degree of carbonation. 

\begin{figure}[H]
\centering
\includegraphics[width=1\linewidth]{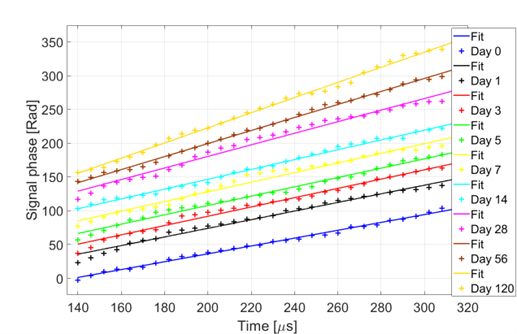}\caption{Signal phase for the 0.6 w/c ratio cement paste specimens and their linear fitting. The numbers shown in the legend are the days exposed to carbonation.}\label{fig7}
\end{figure}

Figure \ref{fig7} shows the variation of the slope of the fitted phase vs. carbonation time data. Only the w/c = 0.6 specimen is presented as an example, but all w/c ratios showed a similar behavior. All specimens showed a linear behavior in terms of phase vs. carbonation time, with correlation coefficients of $>0.98$ for all carbonation durations studied. In some cases, it was not easy to observe any change between different carbonation days. However, if the slope obtained from the linear fitting of each day is plotted against carbonation days, the description of the behavior of the specimens according to their w/c ratio becomes more evident.  

\begin{figure}[H]
\centering
\includegraphics[width=0.9\linewidth]{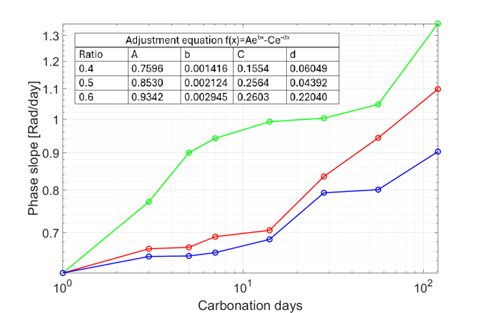}\caption{Exponentially decreasing slope of the signal phases as a function of carbonation days. An anomalous datum was modified, corresponding to day 1 of the 0.4 w/c ratio specimen. The color of the lines indicate the w/c ratio: blue (w/c = 0.4), red (w/c = 0.5), and green (w/c = 0.6).}\label{fig8}
\end{figure}

Figure \ref{fig8} shows that the slopes of the signal phase vs. time plots decreased as the number of days exposed to carbonation increased. The slopes were plotted against the carbonation days, and the resulting curves were fitted to an exponential decay curve with form $A-Be^{-cx}$. It was observed that at first, the slopes abruptly changed. The signal phase of the 0.6 w/c ratio paste changed faster at the beginning of the experiment compared with the other pastes. This was most likely because carbonation occurred faster in this paste, so the signal had more interaction with the products generated early in the process. In the last days of carbonation, the curve corresponding to the 0.6 w/c ratio was constant and presented a higher value than the other curves. A constant slope meant that carbonation occurred at a constant velocity, a process that started first in the 0.6 w/c ratio paste, and then in the 0.5 and 0.4 w/c ratio pastes. Analyzing the equation obtained from the fitting, the decay constants were c = 0.53, 0.83, and 2.03, which showed that the rate of change of the slope with respect to the carbonation days was greater for the 0.6 w/c ratio specimens. In all cases, the coefficients of determination were  $>0.99$.

\begin{figure}[H]
\centering
\includegraphics[width=0.9\linewidth]{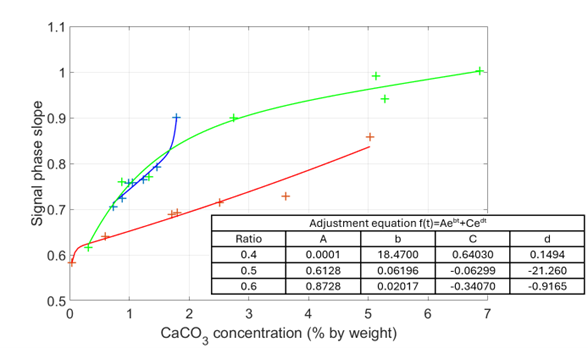}\caption{Phase slope vs. $CaCO_3$ concentration. The colors of the lines indicate w/c ratio: blue (w/c = 0.40), red (w/c = 0.5), and green (w/c = 0.6).}\label{fig9}
\end{figure}

To conclude this linear analysis, the relationship between signal phase slope and $CaCO_3$ concentration is shown in Figure \ref{fig9}. It can be seen that the slope of the phases for the 0.6 and 0.5 w/c specimens changed more slowly than that of the 0.4 w/c specimens. This results provides a direct relationship between the phase behavior of a signal and the degree of carbonation of the sample. To quantify this change, we again fit an exponential of the form $Ae^{bx} + Ce^{dx}$ and obtained the following coefficients of determination (0.93, 0.99, and 0.97). Each of the two exponentials employed for the description of the signal phase slope can be associated with different processes that cause the signal change. The dominant decay coefficients of each w/c ratio were $-3.85$, $-2.04$, and $-1.12$, showing that the change was slower for the 0.6 w/c ratio specimens.

\begin{figure}[H]
\centering
\includegraphics[width=1\linewidth]{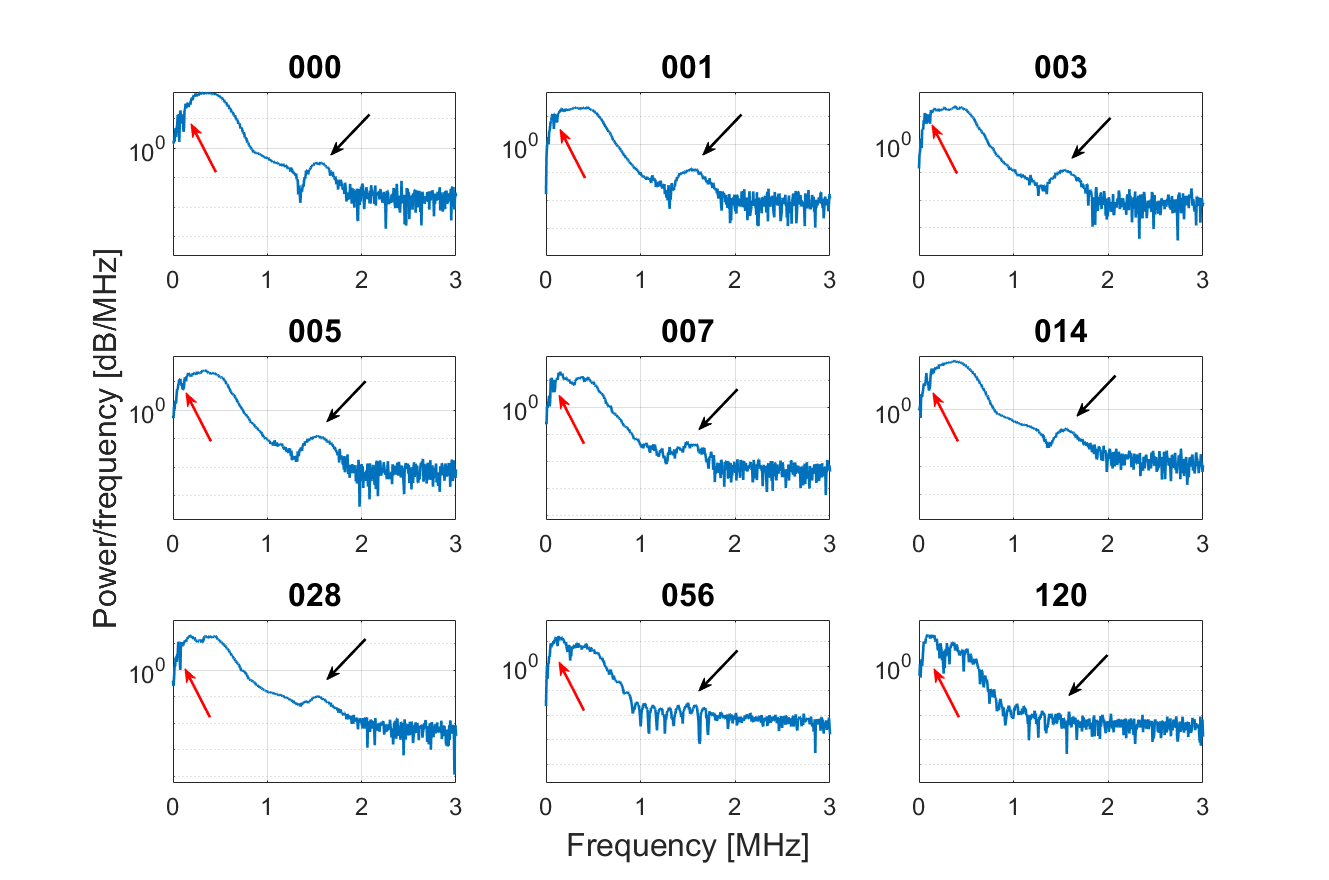}\caption{Power spectral densities for ultrasonic signals obtained for w/c ratio = 0.6 specimens for different carbonation days, as indicated at the tops of the graphs. Similar behaviors were observed for the other w/c ratios.}\label{fig10}
\end{figure}

These results are complemented by applying a non-linear analysis to the signals. As the carbonation process is formed by complex interactions, it is expected that the description would be more adequate in a non-linear regime. For this type of analysis, the power spectral density was obtained for each signal, as shown in Figure \ref{fig10}. The generated harmonics in the signal are due to the non-linear interactions of the induced wave with the media. For this reason, it is expected that this non-linear analysis adequately models the interactions of the wave with the cementitious matrix and carbonation products. 

Figure 10 shows the power spectra of ultrasonic signals at different carbonation days. A fundamental harmonic and a third harmonic (with respect to the fundamental frequency of the transducer) were observed in almost all days. When carbonation was present, these harmonics showed a small shift at low frequencies for all w/c ratio pastes. At 56 and 120 days, the third harmonic was no longer clearly visible in any of the w/c ratio pastes (black arrow). A summary of the frequencies of these harmonics before and after carbonation is presented in Table \ref{tab1}.

\begin{table*}
\begin{centering}\caption{Harmonic frequencies before and after carbonation for all water/cement ratio pastes expressed in kHz. In all cases, a similar standard deviation is obtained.}\vspace{5px}
\resizebox{0.7\textwidth}{!}{\begin{tabular}{|c|c|c|c|c|c|c|}
\hline 
\multirow{2}{*}{ratio} & \multicolumn{2}{c|}{Without carbonation} & \multicolumn{2}{c|}{Carbonated probes (28 day)} & \multicolumn{2}{c|}{Carbonated probes (120 day)}\tabularnewline
\cline{2-7} \cline{3-7} \cline{4-7} \cline{5-7} \cline{6-7} \cline{7-7} 
 & First & Third & First & Third & First & Third\tabularnewline
\hline 
0.4 & 420$\pm$10 & 1530$\pm$20 & 370$\pm$10 & 1510$\pm$20 & 210$\pm$10 & --\tabularnewline
\hline 
0.5 & 390$\pm$12 & 1560$\pm$15 & 360$\pm$12 & 1550$\pm$15 & 340$\pm$12 & --\tabularnewline
\hline 
0.6 & 340$\pm$15 & 1560$\pm$25 & 334$\pm$15 & 1530$\pm$25 & 320$\pm$15 & --\tabularnewline
\hline 
\end{tabular}}\label{tab1}
\par\end{centering}
\end{table*}

We observed that the fundamental harmonic tended to shift towards lower frequencies as the degree of carbonation increased. However, this tendency was inconclusive because some of these shifts were within the uncertainty interval. The frequency variations of the third harmonic were small with respect to the carbonation exposure time but were more significant compared to those of the fundamental, as shown in Figure \ref{fig10}, resulting in a more evident trend to correlate the variation with the carbonation process. 

As previously mentioned, the presence of harmonics is due to the non-linear behavior of the cement paste when interacting with the wave—a behavior that is directly related to porosity \cite{Hodgskinson,Renaud}. We observed that the magnitude of the third harmonic decreased as carbonation progressed, which can assist in the interpretation of the microstructural change in the paste due to carbonation. A possible explanation is that the intensity of this third harmonic is more sensitive to changes in the structure of the paste when the pores contain water or air, rather than $CaCO_3$. Additionally, a subharmonic was observed in all plots (red arrow), which is related to pre-existing cracks in the paste, possibly produced during the drying of the samples \cite{Solodov}. This subharmonic increased in bandwidth and magnitude across all spectra as the carbonation increased, as a result of structural changes due to $CaCO_3$ present in the cracks. Initial fractures were present along the specimen at the beginning of carbonation, and as carbonation progressed, the extension decreased due to the precipitation of $CaCO_3$ filling in those microcracks. 

\begin{figure}[H]
\centering
\includegraphics[width=0.9\linewidth]{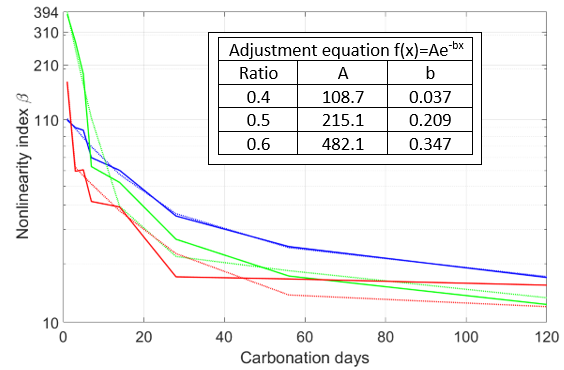}\caption{Non-linearity parameter for the w/c ratio specimens 0.4 (blue), 0.5 (red), and 0.6 (green) for different days of exposure to carbonation.}\label{fig11}
\end{figure}

The non-linearity parameter described in Equation \ref{eq5} \cite{Zhao} can be employed for the evaluation of the change in the third harmonic as carbonation progressed. As shown in Figure \ref{fig11}, the non-linear parameter decreased in all w/c ratio pastes as the number of carbonation days increased, in agreement with decreasing harmonics amplitude. Upon analyzing the fitted exponentials of the form $Ae^{bx}$, it was observed that the coefficient b increased as the w/c ratio increased. The determination coefficient of this relationship was $>0.92$ for all cases. 

\begin{figure}[H]
\centering
\includegraphics[width=0.9\linewidth]{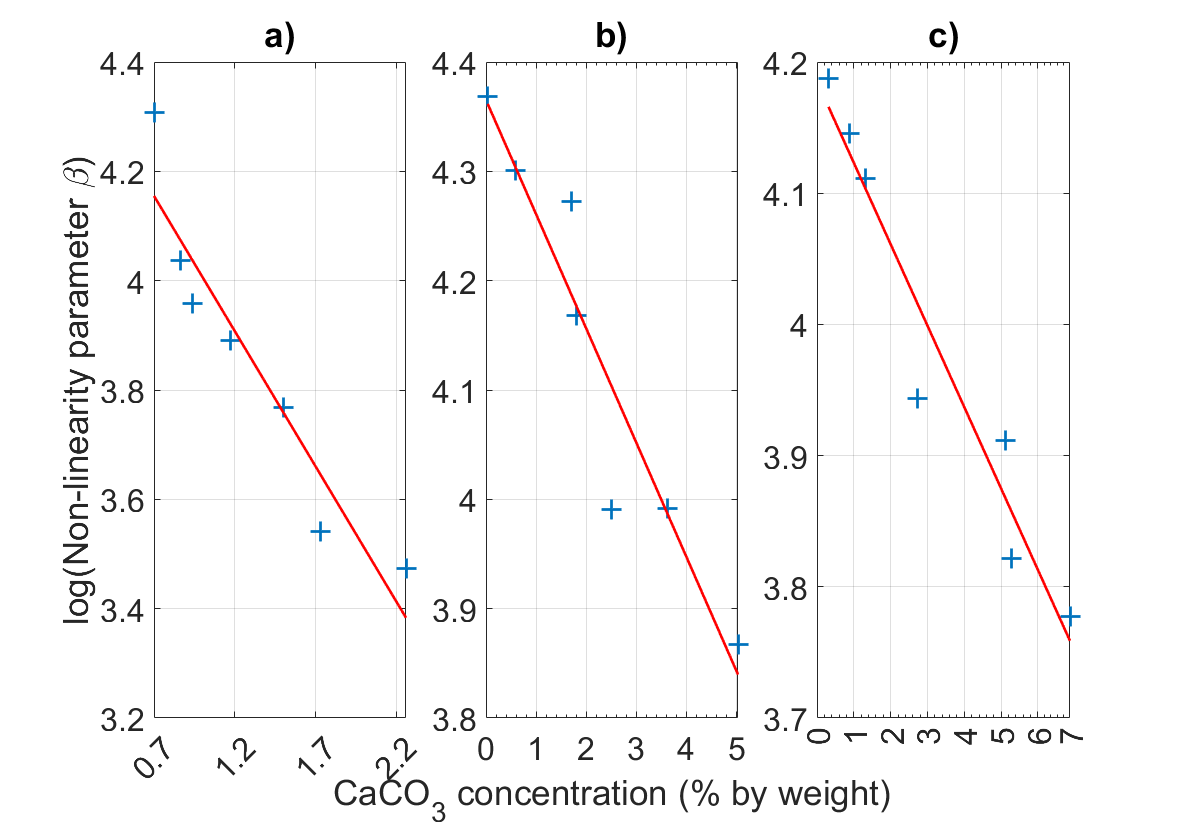}\caption{Non-linearity parameter with respect to $CaCO_3$ concentration, for w/c ratios of a) 0.4, b) 0.5, and c) 0.6.}\label{fig12}
\end{figure}

Figure \ref{fig12} shows the logarithm of the non-linearity parameter as a function of $CaCO_3$ concentration. This logarithm produces decreasing linear relationships with high correlation indexes ($-0.9477$, $-0.9508$, and $-0.9698$) for pastes with w/c ratios of 0.4, 0.5, and 0.6, respectively) at a significance level lower than 0.05. The observed slopes for each w/c ratio were $-0.43$, $-0.10$, and $-0.06$, respectively. In this case, the slope indicated how the non-linearity parameter decreased as the concentration of $CaCO_3$ increased. This change was more relevant for lower w/c ratio samples, which would indicate that the non-linearity parameter decreased more slowly for 0.6 w/c ratio samples than for 0.4 w/c ratio samples with increasing $CaCO_3$ content. This was most likely due to the high porosity of the samples with a high w/c ratio, as the carbonation products were not sufficient in quantity to fill all the pores. However, there was a strong dependance in all w/c ratios, which can indicate the carbonation degree in the pastes in an indirect and non-destructive way.

\section{Conclusions}
As mentioned, it is critical to have a correct carbonation diagnostic method for cement pastes in order to know the integrity of its structure. Although there are many such methods, our results showed that the proposed analysis based solely on acoustic signals might be a benchmark for the diagnosis of carbonation in cement pastes. 

Linear (time and frequency domain) and non-linear (frequency domain) analyses of ultrasonic signals from carbonated cement pastes produced acceptable indices for carbonation in cement pastes with different w/c ratios. These indices can be used to determine the carbonation degree of a cement paste in a non-destructive way, as well as provide a description of carbonation over time. 

Among the obtained indices, the following turned out to be well-correlated to carbonation (determination and correlations coefficients from fitting $>0.89$ and abs($-0.94$), respectively):
\begin{enumerate}
\item The exponential decreasing rate $\mu$ of the normalized cumulative energy varied inversely proportionally with respect to days of exposure to carbonation (Figure \ref{fig3}). 

\item The coefficient of variation of $\Delta t$, depending on the $CaCO_3$ concentration (Figure \ref{fig6}), was directly proportional to the w/c ratio of the paste, and was associated to the carbonation degree of the specimen. 

\item The exponentially decreasing ratio of the slope of the signal phases with carbonation days (Figure \ref{fig8}) increased proportionally with the w/c ratio of the pastes.  

\item In double exponential fitting of the phase slopes in function of $CaCO_3$ concentration (Figure \ref{fig9}), a relationship between the decreasing ratio of the initial exponential of the non-linear parameter and the w/c ratio of the pastes was observed. 

\item Regarding the proposed non-linear analysis in the frequency domain, from the proposed exponential fitting of the non-linearity parameter as a function of carbonation days (Figure \ref{fig11}), the initial intensity and the exponential decreasing ratio were related to the w/c ratio of the pastes. 

\item The slope of the linear fitting of the logarithm of the non-linearity parameter as a function of $CaCO_3$ concentration was associated with the w/c ratio of the pastes (Figure \ref{fig12}). 
\end{enumerate}
The indices obtained from the processing of the ultrasonic signals, with high levels of correlation and determination of the described phenomenon, allow for a description of the carbonation process for the different w/c ratios considered. This description considers both the evolution in time and the degree of carbonation of the sample. 

Further research should evaluate the statistics reliability of these results, along with other indices found in the literature (e.g., those found by Villarreal et al. (2019)\cite{Villarreal} and by Cosmes-López et al. (2017)\cite{Cosmes}), with the final purpose of obtaining a reliable methodology based on indices solely obtained by ultrasonic signals as a more efficient and economical methodology than the traditional ones for the monitoring of carbonation in cement-based materials.

\section*{Declaration of competing interest}
The authors declare that they have no known competing financial interests or personal relationships that could have appeared to influence the work reported in this paper.

\section*{Data availability}
Data will be made available on request.

\section*{Funding}
This research did not receive any specific grant from funding agencies in the public, commercial, or not-for-profit sectors.

\section*{Acknowledgement}
A. Villarreal thanks conacyt for the stipend received and thanks the IPN-CIIDIR for the facilities provided.

\end{document}